\documentclass[aps,prl,twocolumn,amsmath,amssymb,superscriptaddress,floatfix]{revtex4-1}

\usepackage{graphicx}
\usepackage{multirow}
\usepackage{xcolor}
\usepackage{bm}
\usepackage{amsfonts,amssymb,amsmath}
\usepackage{graphicx,dcolumn,bm,color,braket,slashed}
\usepackage{times} 
\usepackage{comment}
\usepackage{array}
\usepackage{siunitx}
\usepackage{url}
\usepackage[normalem]{ulem}

\renewcommand{\Vec}[1]{\mbox{\boldmath$#1$}}
\def\infinity{\infty}
\def\t#1{\textrm{#1}}
\def\ket#1{|#1\rangle }
\def\bra#1{\langle #1 |}
\def\n{\nonumber \\ }
\def\tensor{\otimes}

\newcommand{\overbar}[1]{\mkern 1.5mu\overline{\mkern-1.5mu#1\mkern-1.5mu}\mkern 1.5mu}

\begin{document}


\title{Circular dichroism of emergent chiral stacking orders in quasi-one-dimensional charge density waves}

\author{Sun-Woo Kim}
\affiliation{Department of Physics and Research Institute for Natural Science, Hanyang University, Seoul 04763, Korea}
\affiliation{Department of Physics, Korea Advanced Institute of Science and Technology, Daejeon 34141, Korea}

\author{Hyun-Jung Kim}
\email{h.kim@fz-juelich.de}
\affiliation{Peter Gr\"unberg Institut and Institute for Advanced Simulation, Forschungszentrum J\"ulich and JARA, D-52425 J\"ulich, Germany}

\author{Sangmo Cheon}
\email{sangmocheon@hanyang.ac.kr}
\affiliation{Department of Physics and Research Institute for Natural Science, Hanyang University, Seoul 04763, Korea}
\affiliation{Institute for High Pressure, Hanyang University, Seoul 04763, Korea}

\author{Tae-Hwan Kim}
\email{taehwan@postech.ac.kr}
\affiliation{Department of Physics, Pohang University of Science and Technology (POSTECH), Pohang 37673, Korea}
\affiliation{MPPHC-CPM, Max Planck POSTECH/Korea Research Initiative, Pohang 37673, Korea}

\date[]{}
\begin{abstract}
Chirality-driven optical properties in charge density waves are of fundamental and practical importance.
Here, we investigate the interaction between circularly polarized light and emergent chiral stacking orders in quasi-one-dimensional (quasi-1D) charge-density waves (CDW) with density-functional theory calculations. 
In our specific system, self-assembled In nanowires on Si(111) surface, spontaneous mirror symmetry breaking leads to symmetrically distinct four degenerate quasi-1D CDW structures, which exhibit geometrical chirality.
Such geometrical chirality may naturally induce optically active phenomena even when the quasi-1D CDW structures are stacked perpendicular to the CDW chain direction.
Indeed, we find that left- and right-chiral stacking orders show distinct circular dichroism responses while a nonchiral stacking order does no circular dichroism.
Such optical responses are attributed to the existence of glide mirror symmetry of the CDW stacking orders.
Our findings suggest that the CDW chiral stacking orders can lead to diverse active optical phenomena such as chirality-dependent circular dichroism, which can be observed in scanning tunneling luminescence measurements with circularly polarized light.
\end{abstract}

\maketitle

\newpage

Chirality plays a significant role in light-matter interactions at all branches of the natural sciences including chemistry, biology, and physics~\cite{Hyde1996}.
In condensed matter physics, chiral materials can be a basal platform for application in chiroptoelectronics such as circular dichroism spectroscopy~\cite{Greenfield2006}, quantum computations~\cite{Cherson2006}, all-optical magnetic recording~\cite{Stanciu2007}, and valleytronic devices~\cite{Li2020} to name a few.
Interestingly, chirality appears in charge-density waves (CDW) via symmetry breaking with versatile physical properties~\cite{Ishioka2010,Castellan2013,Xu2020,Jiang2021,Shumiya2021,Wang2021,yang_giant_2020,Kim2020}. %
Among them, $1T$-TiSe$_2$ exhibits three-dimensional (3D) chiral stacking orders of two-dimensional (2D) CDW building blocks~\cite{Ishioka2010,Castellan2013} due to the chiroptoelectric effect~\cite{Xu2020}.

Recently, our previous work~\cite{Kim2020} demonstrated the existence of various 2D CDW chiral stacking orders in the self-assembled quasi-one-dimensional In atomic chains on Si(111) surface~\cite{Yeom1999,Bunk1999,cho2001weakly,Gonzalez2006,Wippermann2010,Kim2012,kim2013driving,zhang2014stabilization,kim2015equivalence,Cheon2015,kim2016origin,Kim2017}.
Here, each In atomic chain is dimerized parallel to the chain generating four degenerate chiral CDW building blocks due to the spontaneous mirror symmetry breaking~\cite{Cheon2015}.
Furthermore, perpendicular to the In atomic chains, the interchain coupling forces the three different stackings of the CDW building blocks to be stabilized, i.e., left-, right-, and nonchiral stacking orders~\cite{Kim2020}.
%
%
Although light-matter interaction in this unique system has been widely studied~\cite{Pedreschi1998,Chandola2009,Speiser2016,Frigge2017,Nicholson2018,Horstmann2020}, such chirality-driven optical properties (such as circular photogalvanic effect~\cite{Xu2020} and circular dichroism~\cite{Greenfield2006}) has not been explored yet.
%


In this Letter, we investigate the light-matter interaction between circularly polarized light and
 CDW stacking orders, using the density-functional theory (DFT) calculations.
We find that left- and right-chiral stacking orders show the circular dichroism, while nonchiral stacking order does not.
We show that the circular dichroism results from the emergent chirality of the chiral stacking orders rather than inherent chirality of a single CDW building block.
Our finding suggests that the CDW chiral stacking orders will pave the practical way to provide circular polarized light emission depending on chirality via electroluminescence.
Based on our results, we propose that a large-scale single domain with the specific chiral stacking order can be selectively induced by utilizing the circularly polarized light.

\begin{figure*}[hbt]
\includegraphics[width=.8\linewidth]{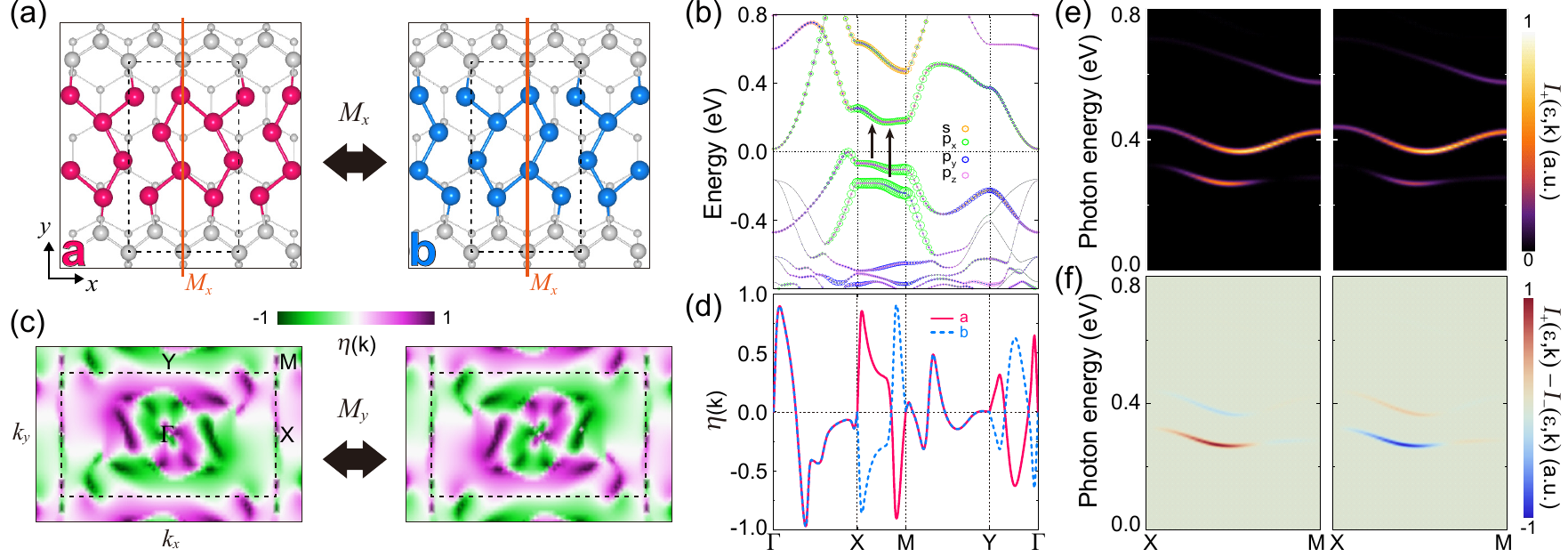} 
\caption{
(a) Atomic structures of degenerate $4\times2$ CDW structures for $a$ and $b$. 
The mirror operation $M_x$ connects two structures.
In atoms are represented by blue and red spheres.
Si atoms in the zigzag chains (the substrate) are represented by the larger (smaller) gray spheres.
Black dashed rectangles indicate unit cells. 
(b) Orbital-projected band structure for $4\times2$ structure. 
Note that $a$ and $b$ have the identical band structures. 
$s$, $p_x$, $p_y$, and $p_z$ orbitals of In atoms are projected on the band structure where the magnitude of open circles is proportional to the projected weight.
(c) 2D maps of degree of circular polarization $\eta_i(\mathbf k) (i=a,b)$ between the highest valence band (HVB) and the lowest conduction band (LCB) for $a$ (left panel) and $b$ (right panel). 
$\eta_a(\mathbf k)$ and $\eta_b(\mathbf k)$ are related by $M_y$.
Black dashed rectangles indicate BZ. 
(d) $\eta_a(\mathbf k)$ and $\eta_b(\mathbf k)$ along the high-symmetry lines.
(e) $k$-resolved interband optical transition rate for left-circularly polarized light $I_{+}(\varepsilon,k)$ and (f) difference between transition rates $I_{+}(\varepsilon,k)-I_{-}(\varepsilon,k)$ for $a$ (left panel) and $b$ (right panel).
}
\end{figure*}


Self-assembled In nanowires on Si(111) consist of two In atomic zigzag subchains that are stitched with adjacent Si chains~\cite{Bunk1999}.
Below $T \approx 125$ K, two indium atomic subchains undergo a CDW transition from the metallic $4\times1$ state to the insulating $4\times2$ state via spontaneous mirror symmetry breaking~\cite{Yeom1999,Gonzalez2006,Wippermann2010,kim2013driving,kim2016origin}.
The broken mirror symmetry leads to the  four energetically-degenerate but symmetrically-distinct $4\times2$ CDW building blocks $\{a,b,c,d\}$, each of which is geometrically chiral.
Among them, chiral and achiral pairs can be assigned depending on their symmetry relations: 
for example, $a$ and $b$ (or $a$ and $d$) are a chiral pair related by the broken mirror reflection symmetry operator $M_x$, while $a$ and $c$ (or equivalently, $b$ and $d$) are an achiral pair related by a half translation along the $x$ direction~\cite{Kim2020}.
As a result, the achiral pair such as $a$ and $c$ shares the same bulk electronic properties. 
On the other hand, the electronic properties between the chiral pair such as $a$ and $b$ are related by $M_x$, as discussed in detail later.

When two kinds of $4\times2$ CDW structures are stacked alternatively along the $y$ direction, 
stacking between the chiral pair (hereafter, we denote as $ab$ and $ad$) is energetically favored over one between the achiral pair ($aa$ and $ac$); see also Fig.~S1 and Table~S1~\cite{Supple}.
Interestingly, this leads to the emergence of chiral ($abcd$ and $badc$) or nonchiral ($ad$) stacking orders, as evidenced by the previous DFT calculation and scanning tunneling microscope (STM) observation~\cite{Kim2020}.

To investigate the effect of the geometrical chirality in CDW on optical properties, we begin to consider $a$ and $b$ structures, which are a chiral pair related by $M_x$ [Fig.~1(a)]. 
$M_x$ gives rise to a symmetry relation between energy bands of $a$ and $b$ structures as $E^{a}_n(k_x,k_y) = E^{b}_n(-k_x,k_y)$, where $n$ is the band index.
Moreover, due to the time-reversal symmetry, the energy bands of each $4\times2$ structure are constrained by
$E^{i}_n(k_x,k_y) = E^{i}_n(-k_x,-k_y)$ where $i=a,b$, and thus $E^{a}_n(k_x,k_y) = E^{b}_n(k_x,-k_y)$.
This is why two $4\times2$ structures, whose Brillouin zones (BZs) are rectangles, show the same energy dispersions and orbital characters
along the high symmetry lines $\overbar{\Gamma X}, \overbar{XM}, \overbar{MY}, \text{and}~ \overbar{Y\Gamma}$ [Fig.~1(b)].
The insulating band structure originated from the CDW formation allows us to explore the role of the chirality of CDW in the optical properties through the direct interband transitions by circularly polarized light.

For optical properties, we calculate the degree of circular polarization $\eta^{nm}(\mathbf k)$~\cite{VASPBERRY}, which is given by
\begin{align}
\begin{split}
 \eta^{nm}(\mathbf k) = \frac{|P_+^{nm}(\mathbf k)|^2-|P_-^{nm}(\mathbf k)|^2}{|P_+^{nm}(\mathbf k)|^2+|P_-^{nm}(\mathbf k)|^2}.
\end{split}
\end{align}
This $\mathbf k$-resolved quantity distinguishes left and right circularly polarized ($\sigma_+$ and $\sigma_-$) light absorption between the valence and the conduction bands with band indices $m$ and $n$, respectively.
Here, $P^{nm}_{\pm}(\mathbf k)=P^{nm}_{x}(\mathbf k)\pm i P^{nm}_{y}(\mathbf k)$, where the matrix element is given by $P^{nm}_{x,y}(\mathbf k)=\bra{\psi_{\mathbf k}^n} \hat p_{x,y} \ket{\psi_{\mathbf k}^m}$.
Berry curvature of the $n$-th band in terms of the matrix element $P^{nm}_{x,y}(\mathbf k)$ is given by~\cite{yao2008valley,chang1996berry}
\begin{align}
\begin{split} \label{eq:berry_circular}
 \Omega_z^n(\mathbf k) 
 = & -\frac{\hbar^2}{2m_e^2} \sum_{m\neq n} \frac{\eta^{nm}(\mathbf k)[|P_+^{mn}(\mathbf k)|^2+|P_-^{mn}(\mathbf k)|^2] }{\left[E_n(\mathbf k)-E_m(\mathbf k)\right]^2}. 
\end{split}
\end{align}
From this equation, we note that the symmetry transformation property of $\eta^{nm}(\mathbf k)$ follows that of $\Omega_{z}^{n}(\mathbf k)$.
The k-resolved interband optical transition rate between valence and conduction bands is obtained from Fermi’s golden rule within the electric dipole approximation~\cite{fang2015ab}:
\begin{align}
\begin{split} \label{eq:interband_opt}
I_{\pm}(\varepsilon,\mathbf k) \sim \frac{\hbar^2}{\varepsilon^2} \sum_{n,m} |P^{nm}_{\pm}(\mathbf k)|^2 \delta(E_n(\mathbf k)-E_m(\mathbf k)-\varepsilon).
\end{split}
\end{align}
We only consider the direct optical transition at $\mathbf k$ between valence and conduction bands. The $1/\varepsilon^{2}$ factor comes from the conversion between the square of the gauge field $A^2$ and the incoming light intensity $E^2$.

The symmetry relation between $a$ and $b$ structures also reveals in the degree of circular polarization $\eta^{nm} (\mathbf k)$ of each structure, hereafter $\eta_{i} (\mathbf k)$ ($i=a,b$) [Fig.~1(c)].
Due to the absence of inversion and mirror symmetries, $\eta_i(\mathbf k)$ shows nonzero finite values and $\eta_{i}(k_x,k_y)$ is odd owing to time-reversal symmetry $\Theta$: $\eta_{i}(k_x,k_y) = - \eta_{i}(-k_x,-k_y)$.
In addition, $M_x$ gives rise to the following relation between two structures as $\eta_{a}(k_x,k_y) =  - \eta_{b}(-k_x,k_y)$.
%
Hence, the combination of $\Theta$ and $M_x$ leads to  
$\eta_{a}(k_x,k_y) = \eta_{b}(k_x,-k_y)$, which means that they are related by $M_y$. 
The relation between $\eta_{a}(\mathbf k)$ and $\eta_{b}(\mathbf k)$ is clearly seen in the line profiles along the high-symmetry lines [Fig.~1(d)].
They are the same along the $\overbar{\Gamma X}$ and $\overbar{MY}$ lines 
while they are opposite along the $\overbar{XM}$ and $\overbar{Y\Gamma}$ lines, showing a strong momentum-selective circular dichroism of $a$ and $b$ structures.

In order to see the optical transition rate under left (right) circularly polarized light $\sigma_+ (\sigma_-)$, which is more appropriate quantity that can be measured in optical experiments,
we calculate the $\mathbf k$-resolved interband optical transition rate for $I_{\pm}(\varepsilon,\mathbf k)$ in Eq.~(\ref{eq:interband_opt}) along the high-symmetry lines $\overbar{\Gamma X}, \overbar{XM}, \overbar{MY}, \text{and}~ \overbar{Y\Gamma}$.
We find that $I_{\pm}(\varepsilon,\mathbf k)$ has nonzero finite values only along the $\overbar{XM}$ line [Fig.~1(e)] where the $p_x$ orbital character is strong [Fig.~1(b)].
Furthermore, we find that the circular dichroism $I_{+}(\varepsilon,\mathbf k)-I_{-}(\varepsilon,\mathbf k) $ of the $a$ structure along the $\overbar{XM}$ line is opposite to its chiral partner $b$ [Fig.~1(f)], which is consistent with $\eta_{i}(\mathbf k)$ [Fig.~1(d)].
Note that the magnitude of the circular dichroism can be enhanced by manipulating the degree of symmetry breaking (see Fig.~S2~\cite{Supple}).


\begin{figure}
\includegraphics[width=0.8\linewidth]{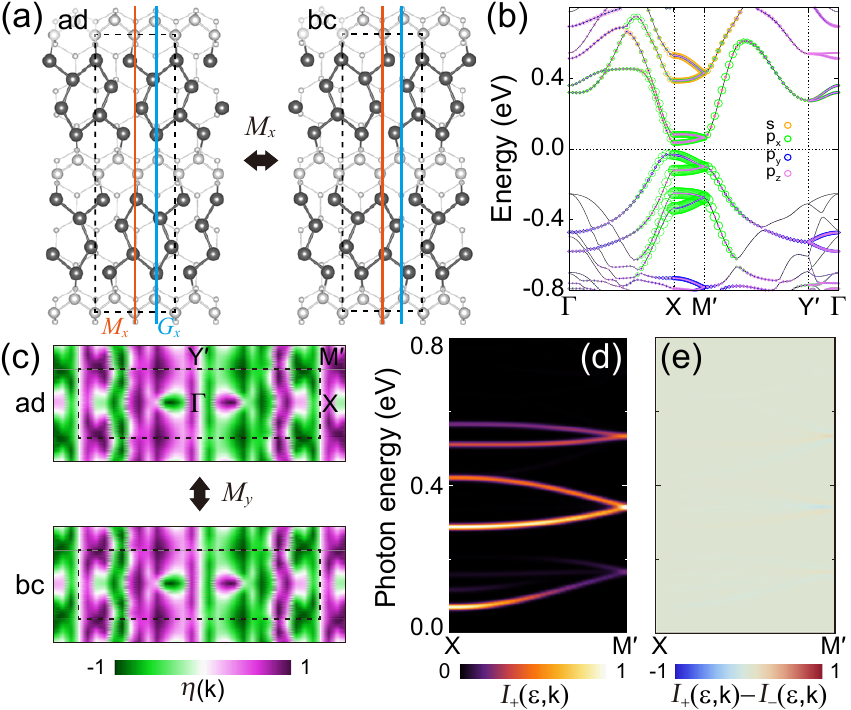} 
\caption{
(a) Atomic structures of nonchiral stacking orders $ad$ and $bc$, which are related by $M_x$.
Black dashed rectangles indicate unit cells.
There is the glide mirror symmetry $G_x$ for nonchiral stacking orders.
(b) Orbital-projected band structure for the nonchiral stacking order ($ad$ and $bc$ have identical band structures).
(c) 2D map of $\eta_i(\mathbf k)$ ($i=ad, bc$) between HVB and LCB for $ad$ (upper panel) and $bc$ (lower panel), which are related by $M_y$.
Black dashed rectangles indicate BZ. 
(d) $I_{+}(\varepsilon,k)$ and (e) $I_{+}(\varepsilon,k)-I_{-}(\varepsilon,k)$ for the nonchiral stacking order.
}
\end{figure}

\begin{figure*}[hbt]
\includegraphics[width=.8\linewidth]{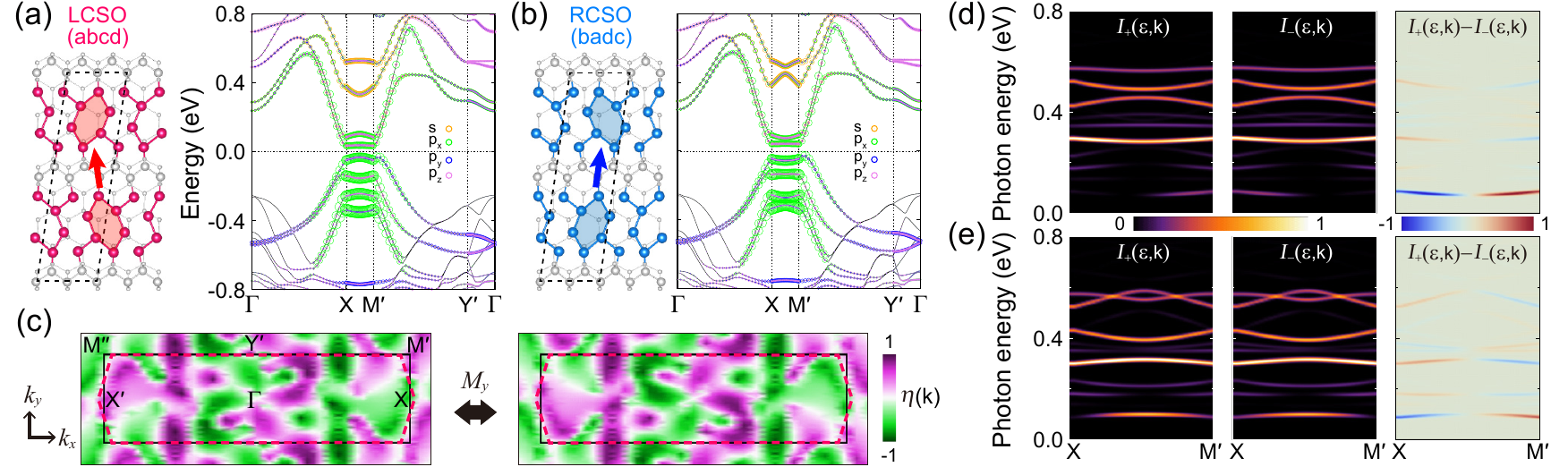} 
\caption{
(a),(b) Atomic structures and orbital-projected band structures of (a) left-chiral ($abcd$) and (b) right-chiral ($badc$) stacking orders.
Two chiral stacking orders are related by $M_x$.
Black dashed parallelograms indicate unit cells.
(c) 2D maps of $\eta_i(\mathbf k)$ between HVB and LCB for the left-chiral (left panel) and right-chiral (right panel), which are related by $M_y$.
Here, red dashed hexagons indicate BZ of the chiral stacking orders while black rectangles indicate BZ of the nonchiral stacking orders in Fig.~2. 
(d),(e) $I_{+}(\varepsilon,k)$, $I_{-}(\varepsilon,k)$, and $I_{+}(\varepsilon,k)-I_{-}(\varepsilon,k)$ for the (d) left-chiral and (e) right-chiral stacking orders.
}
\end{figure*}

Based on the understanding of the optical response of $4\times2$ CDW structures,
we investigate the nonchiral stacking orders (NCSO) composed of two $4\times2$ CDW structures.
We consider two representative NCSO $ad$ and $bc$, which are related by $M_x$ [Fig.~2(a)].
$M_x$ gives rise to the symmetry relation on the energy bands of the two as $E^{ad}_n(k_x,k_y) = E^{bc}_n(-k_x,k_y)$.
Moreover, due to the glide mirror symmetry $G_x$, $E^{i}_n(k_x,k_y) = E^{i}_n(-k_x,k_y)$ holds and thus we obtain $E^{ad}_n(\mathbf k) = E^{bc}_n(\mathbf k)$.
Hence, their band structures are the same along the lines $\overbar{\Gamma X}, \overbar{XM'}, \overbar{M'Y'}, \text{and}~ \overbar{Y'\Gamma}$ [Fig.~2(b)].
The orbital-projected band structure shows the direct band gap at the $X$ point with the strong $p_x$ orbital character near the Fermi level.

For the optical properties, $\eta_{ad}(\mathbf k)$ and $\eta_{bc}(\mathbf k)$ are related by $M_y$ from $M_x$ and $\Theta$ [Fig.~2(c)]. 
Notably, owing to the presence of the glide mirror symmetry $G_x$ [Fig.~2(a)] as well as $\Theta$, $\eta_i(\mathbf k)$ of each NCSO has the symmetry constraint as $\eta_i(k_x,k_y)=\eta_i(k_x,-k_y)$, which means that $\eta_i(\mathbf k)$ of each NCSO is itself $M_y$ symmetric.
Hence, with the relation $\eta_{ad}(k_x,k_y) = \eta_{bc}(k_x,-k_y)$ constrained by $M_x\Theta$,  we obtain $\eta_{ad}(\mathbf k)=\eta_{bc}(\mathbf k)$ [Fig.~2(c)].
This leads to the same optical transition rates $I_{\pm}(\varepsilon,\mathbf k)$ between $ad$ and $bc$ [Fig.~2(d)].
Similar to the $a$ and $b$ structures, there is no $I_{\pm}(\varepsilon,\mathbf k)$ intensity except for the $\overbar{XM'}$ line where $p_x$ orbital character is strong.
Moreover, unlike the $a$ and $b$ structures, the difference $I_{+}(\varepsilon,\mathbf k)-I_{-}(\varepsilon,\mathbf k)$ [Fig.~2(e)] shows no circular dichroism for NCSO.
This is consistent with the intuition that the nonchiral stacking order does not exhibit circular dichroism due to the cancellation of the opposite chiralities between chiral partners $a$ and $d$ (or $b$ and $c$) when stacked along the $y$ direction.

We now consider chiral stacking orders composed of four subsequent $4\times2$ CDW structures, which are classified as left- and right-chiral stacking orders (LCSO and RCSO) represented by $abcd$ and $badc$, respectively [Figs.~3(a) and 3(b)].
Here, the unit cell of chiral stacking orders is chosen as a parallelogram with the $8\times2$ periodicity [dashed lines in Figs.~3(a) and 3(b)], 
which is different from the rectangle unit cell with the $16\times2$ periodicity of the previous study~\cite{Kim2020}. 
Since the two chiral stacking orders are related by $M_x$, they are the chiral partner of each other.
$M_x$ and $M_x\Theta$ impose constraints on the energy bands of two different chiral stacking orders,
$E^{abcd}_n(k_x,k_y) = E^{badc}_n(-k_x,k_y)$
and $E^{abcd}_n(k_x,k_y) = E^{badc}_n(k_x,-k_y)$, respectively.
Thus, LCSO and RCSO have identical band structures along the $\overbar{Y'\Gamma}$ $(k_x=0)$ and $\overbar{\Gamma X}$ $(k_y=0)$ lines [Figs.~3(a) and 3(b)].
However, their band structures are different along the $\overbar{XM'} (k_x=\pi)$ and $\overbar{M'Y'} (k_y=\pi)$ lines, in contrast to the above mentioned $4\times2$ and NCSO cases. 
The difference between chiral and nonchiral stacking orders results from different shapes of BZs: LCSO and RCSO have the distorted hexagonal BZ [red lines in Fig.~3(c)] while the $4\times2$ and NCSO do the rectangular BZ [Figs.~1(c) and 2(c)]. 
Note that, due to the $M_x$, band structure of LCSO along the $\overbar{XM'}$ is the same with that of RCSO along the $\overbar{X'M''}$.

For the optical properties, $\eta_{abcd}(\mathbf k)$ and $\eta_{badc}(\mathbf k)$ are also related by $M_y$ from $M_x\Theta$ [Fig.~3(c)].
Due to the $M_x$, $\eta_{abcd}(\mathbf k)$ along the $\overbar{XM'}$ is the opposite of $\eta_{badc}(\mathbf k)$ along the $\overbar{X'M''}$.
The chiral partners $abcd$ and $adcb$ have a difference in the optical transition rates $I_{\pm}(\varepsilon,\mathbf k)$ along the $\overbar{XM'}$ [Figs.~3(d) and 3(e)], which is originated from the different band structures in Figs.~3(a) and 3(b).
This feature is different from the NCSO case where the chiral partners $ad$ and $bc$ share the same $I_{\pm}(\varepsilon,\mathbf k)$.

Even though LCSO and RCSO have the same numbers of chiral partners in the stacking configuration, they exhibit nonzero circular dichroism $I_{+}(\varepsilon,\mathbf k)-I_{-}(\varepsilon,\mathbf k)$ along the $\overbar{XM'}$ in contrast to NCSO.
This striking result implies that the nonvanishing circular dichroism arises from the emergent chirality of two chiral stacking orders rather than from the inherent chirality of the $4\times2$ building blocks themselves.
The emergent chiralities of two chiral stacking orders can be geometrically characterized by using the opposite phase shift vectors [arrows marked in Figs.~3(a) and 3(b)] and topological chiral winding numbers~\cite{Kim2020}.
Note that, despite the opposite emergent chiralities between two chiral stacking orders, circular dichroism of the two in Figs.~3(d) and 3(e) are not exactly opposite to each other due to the $8\times2$ unit cell used~\cite{note}.

\begin{figure}[hb]
\includegraphics[width=.8\linewidth]{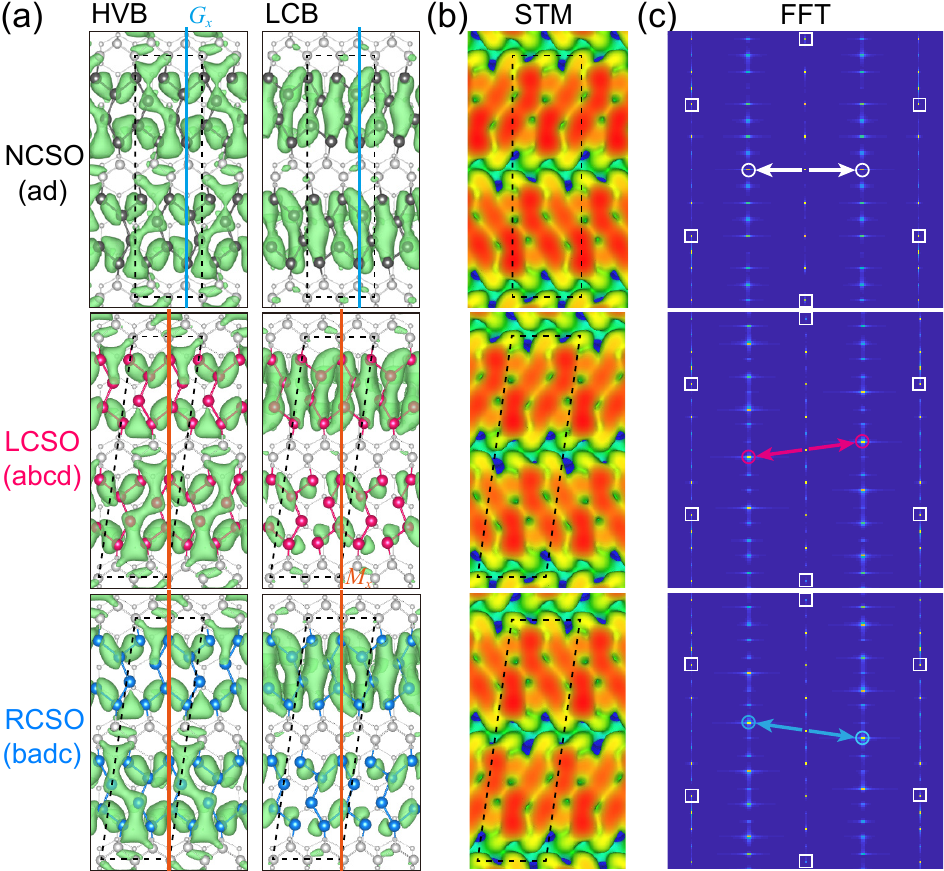} 
\caption{
(a) Calculated charge densities of HVB and LCB states at the $X$ point for NCSO, LCSO, and RCSO.
The charge densities are drawn with an isosurface of $4\times10^{-4}$ electrons$/\text{\AA}^3$.
Dashed parallelograms indicate unit cells.
(b) Simulated filled state STM images (at $-0.3$ V) and (c) corresponding Fast Fourier Transform (FFT) images of three stacking orders~\cite{FFT_details}.
In (c), squares and circles indicate $1 \times 1$ of Si substrates and $8 \times 2$ of stacking orders, respectively.
}
\end{figure}

To visualize the electronic nature of the emergent chirality, we plot the charge densities of the highest valence band (HVB) and the lowest conduction band (LCB) states at the $X$ point for the three stacking orders in Fig.~4(a).
For NCSO $ad$, the charge densities are evenly distributed in two In nanowires separated by Si zigzag chains, keeping the glide symmetry $G_x$ of the system.
The glide symmetry hinders the chirality from emerging and thus there is no circular dichroism [Fig.~2(e)].
However, for the chiral stacking orders $abcd$ and $badc$, the charge densities are unevenly distributed due to the absence of the glide symmetry. 
The absence of the glide symmetry allows the emergence of chirality, thereby leading to nonzero circular dichroism [Figs.~3(d) and 3(e)].
Note that the charge densities of two systems are related by $M_x$ as 
 $|\psi^{abcd}_{(k_x,k_y)}(x,y)|^2 = |\psi^{badc}_{(-k_x,k_y)}(-x,y)|^2$.

Furthermore, the emergent chirality is also seen in the filled state STM images simulated by DFT calculations [Fig.~4(b)] and their Fast Fourier Transform (FFT) images [Fig.~4(c)].
In FFT images, arrows connecting primary FFT peaks (colored by white, red, and blue) clearly highlight the distinct chiralities of three different CDW stacking orders.
We can also observe more or less consistent behaviors in FFT images of experimentally measured filled state STM images depending on chirality (see Fig.~S3~\cite{Supple}).

Thanks to the light-matter interaction, it may be able to grow a single CDW chiral stacking order up to macroscopic scale by cooling down under circularly polarized light.
By switching the handedness of circularly polarized light, one may tune the chirality of stacking orders as demonstrated in $1T$-TiSe$_2$~\cite{Xu2020}.
Once such a macroscopic single chiral stacking order is obtained below the CDW transition temperature, the corresponding circular dichroism response of each chiral stacking order could be experimentally probed.
In addition, the macroscopic single chiral stacking order would exhibit highly-resolved FFT images as shown in the simulated STM images [Fig.~4(c)].

Based on the first-principles DFT calculations, we have investigated the unprecedented optical properties of CDW stacking orders in self-assembled In nanowires on Si(111).
We found that the left- and right-chiral stacking orders show the distinct circular dichroism along the high symmetry lines, while the nonchiral stacking order does not.
The circular dichroism is attributed to the emergent chirality of the chiral stacking orders rather than the inherent chirality of a single CDW building block.
We also revealed that the broken glide mirror symmetry of the chiral stacking orders is responsible for the microscopic origin of the circular dichroism. 
Our findings suggest that the chiral stacking orders of quasi-1D CDW wires are optically active, which can lead to diverse optical active phenomena such as circular dichroism or circularly polarized luminescence depending on chirality.
Additionally, our work will shed light on understanding of light-matter interaction in recent chiral CDW materials such as  AV$_3$Sb$_5$~\cite{Jiang2021,Shumiya2021,Wang2021}.

\begin{acknowledgments}
This work was supported by the National Research Foundation of Korea (NRF) funded by the Ministry of Science and ICT, South Korea (Grants No. 2016K1A4A4A01922028, 2021R1A6A1A10042944,
NRF-2018R1C1B6007607, NRF-2021R1H1A1013517, and  NRF-2021R1F1A1063263).
S.-W.K., H.-J.K., and S.C. acknowledge support from POSCO Science Fellowship of POSCO TJ Park Foundation.
H.-J.K. gratefully acknowledges financial support from the Alexander von Humboldt Foundation.
%
%
We thank the Korea Institute for Advanced Study for providing computing resources (KIAS Center for Advanced Computation Linux Cluster System) for this work.
\end{acknowledgments}

\bibliographystyle{apsrev4-1}

\begin{thebibliography}{38}
\bibitem [{not()}]{note}%
  \BibitemOpen
  \href@noop {} {}\bibinfo {note} {If the $16\times2$ rectangular unit cell is
  used to calculate the circular dichroism, two chiral stacking orders should
  exhibit the opposite circular dichroism in similar to Fig.~1(f) of the
  opposite circular dichroism between $4\times2$ $a$ and $b$
  structures.}\BibitemShut {Stop}%
\bibitem [{FFT()}]{FFT_details}%
  \BibitemOpen
  \href@noop {} {}\bibinfo {note} {We use the unit cell size of $\sim$ 30~nm to
  obtain high-quality FFT images.}\BibitemShut {Stop}%
\end{thebibliography}
%


\clearpage

\pagebreak
\onecolumngrid
\widetext
\renewcommand{\Vec}[1]{\mbox{\boldmath$#1$}}
\def\infinity{\infty}
\def\t#1{\textrm{#1}}
\def\ket#1{|#1\rangle }
\def\bra#1{\langle #1 |}
\def\n{\nonumber \\ }
\def\tensor{\otimes}
\setcounter{figure}{0}
\renewcommand{\theequation}{S\arabic{equation}}
\renewcommand{\thefigure}{S\arabic{figure}}
\renewcommand{\thetable}{S\arabic{table}}
\renewcommand{\thesection}{\normalsize \arabic{section}}
\renewcommand{\thesubsection}{\thesection.\arabic{subsection}}
\renewcommand{\thesubsubsection}{\thesubsection.\arabic{subsubsection}}
    
\begin{center}
\textbf{\LARGE{Supplemental Material: Circular dichroism of emergent chiral stacking orders in quasi-one-dimensional charge density waves
}}
\end{center}


\section{\large Computational details}

We performed density-functional theory (DFT) calculations using Vienna $ab$ $initio$ simulation package (VASP) with the projector-augmented wave method~\cite{VASP1,VASP2,PAW}.
For the exchange-correlation energy, we used the strongly constrained and appropriately normed (SCAN) functional~\cite{SCAN}.
The $\mathbf{k}$-space integration were done with dense 392 and 196 $\mathbf{k}$-points in the surface Brillouin zone of the $4\times2$ and $8\times2$ unit cells, respectively.
The Si(111) substrate below the In wires was modeled by a six-layer slab with $\sim 30$~$\text{\AA}$ of vacuum in between the slabs and the bottom Si layer was passivated by H atoms. 
All the atoms except the bottom two layers were allowed to relax until all the residual force components were less than $0.02$~eV/$\text{\AA}$.  
We note that SCAN functional tends to underestimate the band gap but the overall band structure and the energetics between various $8 \times 2$ insulating states are comparable to the previously reported HSE+vdW results of hybrid exchange-correlation functional~\cite{kim2013driving,zhang2014stabilization,kim2015equivalence,kim2016origin,Kim2020}, as shown in Table~\ref{TableS1}.

\begin{table}[h]
\caption{
\label{TableS1}
Calculated total energies $\Delta E$ (in meV per $8\times2$ unit cell) of various stacking configurations ($aa$, $ab$, $ac$, $ad$, $abcd$, and $badc$)~\cite{Kim2020}
relative to the configuration $ad$ and the band gaps $E_g$ of each structure (in meV).
Note that $ad$, $abcd$, $badc$ correspond to the nonchiral, left-chiral, and right-chiral stacking orders, respectively.
}
{\renewcommand{\arraystretch}{1.7}%
\begin{tabular}{ c c  c   c   c   c  c c }
\hline
\hline
&  & $aa$ & $ab$ & $ac$ & $ad$ & $abcd$ & $badc$ \\
\hline
SCAN (this work) &
$\Delta E$ & 80.9 & 2.7 & 92.0 & 0.0 & -2.6 & -3.6
\\
& $E_g$ & 19 & 39 & 60 & 67 & 67 & 67
\\
HSE+vdW~\cite{Kim2020} &
$\Delta E$ & 84.2 & 4.3 & 106.5 & 0.0 & 0.0 & 0.0
\\
& $E_g$ & 283 & 299 & 305 & 313 & 306 & 306
\\
\hline
\hline
\end{tabular}}
\end{table}

\section{\large Four symmetrically distinct $8\times2$ structures}

\begin{figure}[hbt]
\includegraphics[width=165mm]{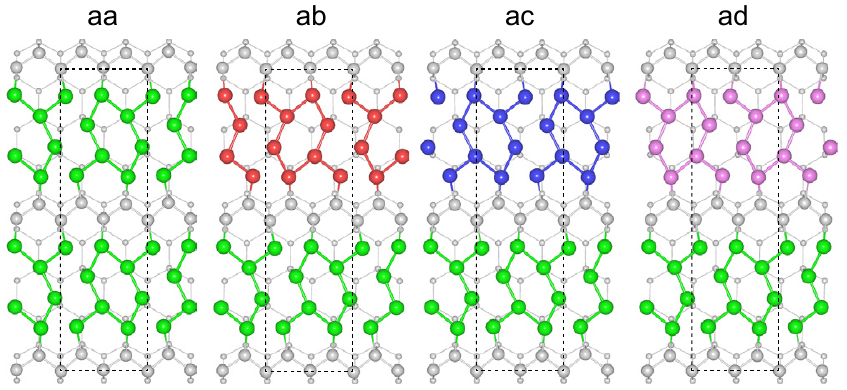} 
\caption{
Four symmetrically distinct $8\times2$ structures ($aa, ab, ac,$ and $ad$)
}
\end{figure}

\newpage

\section{\large Strain dependence of circular dichroism magnitude}

In general, the magnitude of the circular dichroism is proportional to the degree of symmetry breaking that generates chirality. 
In our CDW system, the degree of symmetry breaking is characterized by the bond length difference $\Delta_1-\Delta_2$ as displayed in Fig.~S2(a).
By performing additional DFT calculations, we find that $\Delta_1-\Delta_2$ changes under the strain: the degree of symmetry breaking is enhanced (reduced) under tensile (compressive) strain with lattice constant $a=1.005a_0~(0.995a_0)$.
This is consistent with the previous report~\cite{zhang2014stabilization} that the symmetry broken CDW phase is stabilized (destabilized) over symmetric metallic phase when there is tensile (compressive) strain.
We then check the strain dependence of the circular dichroism magnitude, as shown in Fig.~S2(b).
As expected, the magnitude of circular dichroism follows the same tendency to the degree of symmetry breaking, i.e., tensile strain enhances the circular dichroism magnitude.
In addition, we expect that the circular dichroism magnitude can be enhanced when the hole doping that stabilizes CDW phase~\cite{kim2015equivalence} is introduced.

\begin{figure}[hbt]
\includegraphics[width=0.9\textwidth]{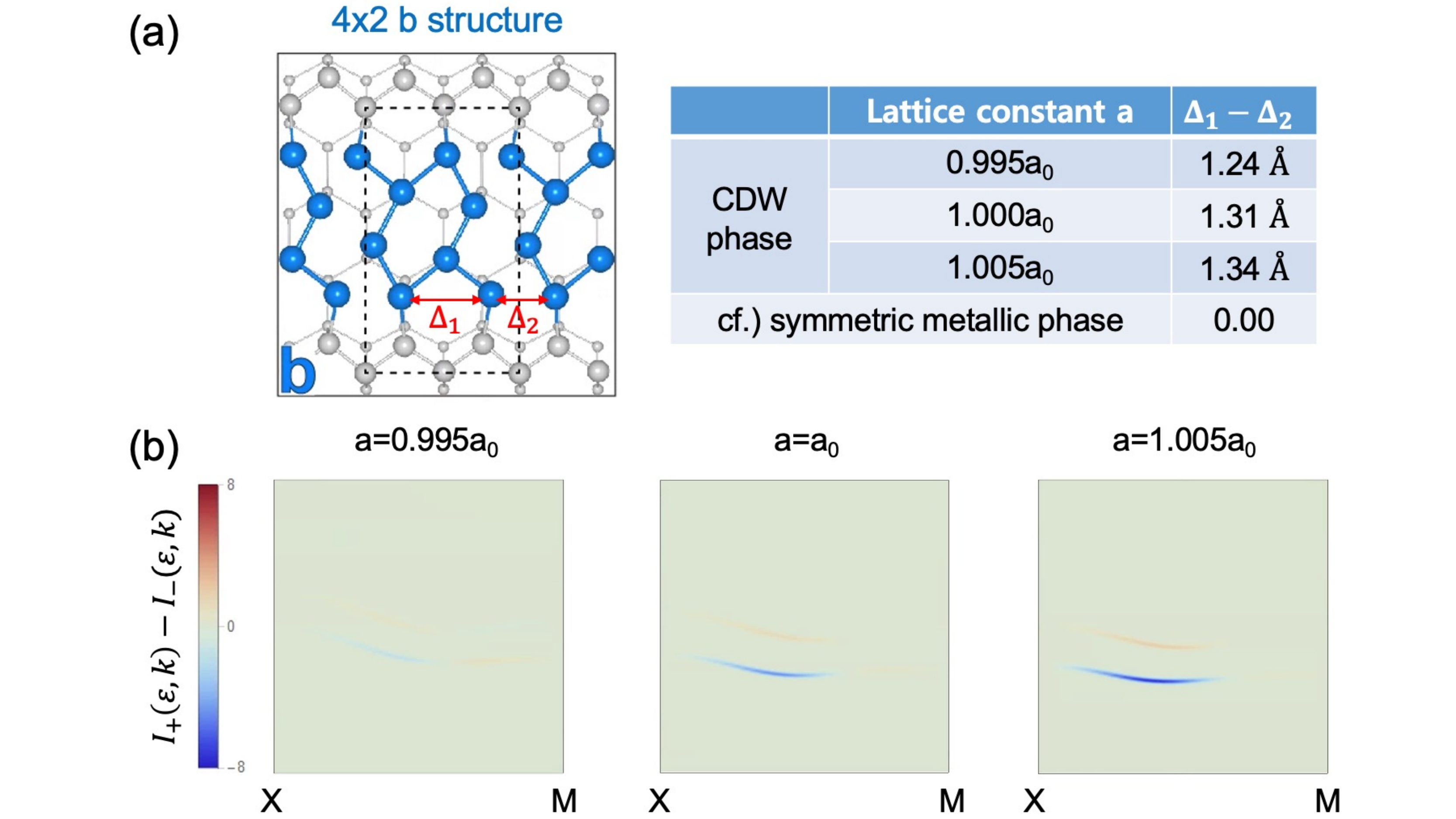} 
\caption{
(a) Geometry and (b) circular dichroism $I_{+}(\varepsilon,k)-I_{-}(\varepsilon,k)$ of the $4\times2$ $b$ structure as a function of a lattice constant $a$.
}
\end{figure}

\clearpage

\section{\large Experimentally measured STM and FFT images}

\begin{figure}[hbt]
\includegraphics[width=120mm]{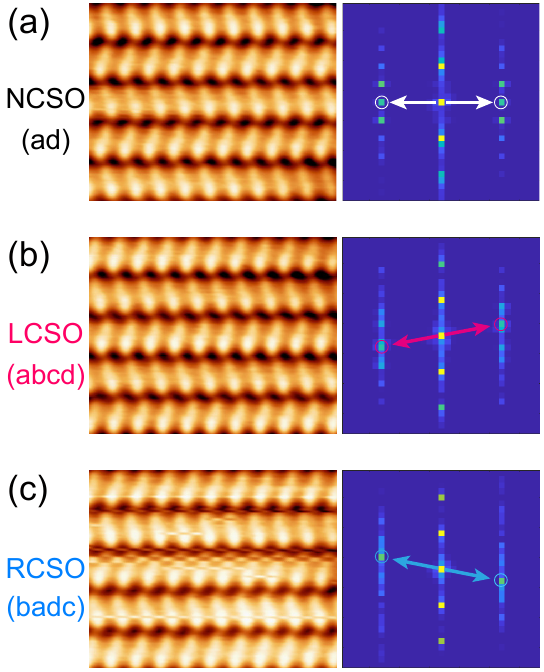}
\caption{
STM images of (a) nonchiral, (b) left-chiral, and (c) right-chiral stacking orders and corresponding FFT images.
All STM images were obtained at $-0.5$~V of sample bias with respect to a tip.
Although the resolution in FFT images are poor due to the small scan size ($\approx 5$~nm) of the STM images, 
we can observe the same chiralities as the larger simulated STM images (Fig.~4 in the main text).
}
\end{figure}

\end{document}